\documentclass[fleqn,twoside]{article}
\usepackage{espcrc2}

\usepackage[dvips]{graphicx}

\title{Electronic band structure of  low-temperature YB$_{12}$, YB$_6$\\
 superconductors and layered YB$_2$ , MgB$_2$ diborides}
\author{I.R. Shein\thanks{E-mail: shein@ihim.uran.ru}, S.V. Okatov, N.I.Medvedeva and A.L.
Ivanovskii}

\begin{document}
\begin{abstract}

Institute of Solid State Chemistry, Ural Branch of the Russian
Academy of Sciences, 620219, Ekaterinburg, Russia\\\

Electronic band structure of boron-rich low-temperature
superconductors  UB$_{12}$-like dodecaboride YB$_{12}$ and
CaB$_6$-like hexaboride YB$_6$ are investigated using the
first-principle FLMTO calculations and compared with one for
layered YB$_2$ and the new "medium-T$_c$" superconductor MgB$_2$
diborides.\\



\end{abstract}

\maketitle

\section{Introduction}
The discovery of the superconductivity in MgB$_2$ (T$_C$ $\approx$
40 K) \cite{Nagamatsu} and creation of promising materials based
thereon (in the form of single crystals, ceramics, thin films,
tapes and wires, see reviews \cite{Ivanovskii,Buzea} have
attracted a great deal of interest in superconducting properties
of other boron-containing phases.\\
Comparison between different classes of binary (semi-(M$_2$B),
mono-(MB), di- (MB$_2$), tetra-(MB$_4$)) and some highest borides
(hexa-(MB$_6$), dodeca-(MB$_{12}$) and MB$_{66}$-like borides),
ternary and quaternary borides (review \cite{Buzea}) shows that
the majority of known superconductors (SC) are found among
low-boron-containing compounds (B/M $\leq$ 2 - 2.5), in which B
atoms are in the form of isolated groups (atoms) or planar
sublattices (boron sheets). The superconducting state is far less
typical of highest borides (B/M $\geq$ 6) having a structure made
up of stable polyhedra of boron atoms (octahedra B$_6$ (MB$_6$),
icosahedra B$_{12}$ (MB$_{12}$) or their combination (MB$_{66}$)).
Among a large number of these boron-rich phases, the
low-temperature superconductivity was observed only for eight
compounds: MB$_6$ (M = Y, La, Th, Nd) and MB$_{12}$ (M = Sc, Y,
Zr, Lu) \cite{Buzea}.\\
It is significant that (i) lower borides of these metals, in
particular Sc and Y diborides, are not SC, and (ii) the stable
polymorphy of the elemental boron ($\alpha$ -B$_{12}$,
$\beta$-B$_{105}$), which under equilibrium conditions contain the
boron polyhedra (icosahedra or "gigantic" icosahedra B$_{84}$) as
structural elements, are semiconductors [4-8]. Only recently it
was found that polycrystalline boron (rombohedral -B$_{105}$)
transforms from a semiconductor to a superconductor (T$_c$
$\approx$ 11.2 K) at about 250 GPa \cite{Eremets}. In this work we
calculate the electronic band structures of low-temperature SCs -
boron-rich phases YB$_{12}$ and YB$_6$ - and compare them with two
representatives of layered AlB$_2$-type diborides, namely the
non-superconducting YB$_2$ and new "medium-T$_c$" SC MgB$_2$. The
results obtained are analyzed in terms of (i) electronic bands,
(ii) density of states (DOS) and (iii) site-projected l-decomposed
DOS near the Fermi energy (E$_F$) of these borides. The Fermi
surfaces for YB$_2$ and MgB$_2$ are also presented.

\section{Structures and computational}
The basic structural elements of the cubic dodecaboride YB$_{12}$
are stable polyatomic boron clusters with icosahedral symmetry
(B$_{12}$) similar to that in elemental boron. The structure of
the UB$_{12}$ type (space group is O$^5$$_h$-Fm3m) is formally
described in terms of simple rock-salt lattice, where Y occupies
Na sites and B$_{12}$ icosahedra are located in Cl sites, the unit
cell contains 52 atoms (Z = 4). The atomic positions are  4M (a)
0,0,0; 48B (i) $\frac{1}{2}$, x, x (x = 0.17011 for our
self-consistent calculation).\\ The role of Y and B$_{12}$ in
forming the YB$_{12}$ band structure can be elucidated by removing
Y atoms from the lattice entirely and calculating this
hypothetical "dodecaboride" with an empty Y-sublattice (EB$_{12}$
- E - structure vacancy) and the icosahedral phase (BB$_{12}$),
which is a result of the Y$\rightarrow$ B substitution.\\ Yttrium
hexaboride has a CaB$_6$-type structure (space group is
O$^1$$_h$-Pm3m). It can be formally described in terms of a simple
CsCl lattice, where metal atoms occupy Cs sites, while B$_6$
octahedra are in Cl sites. The unit cell contains 7 atoms (Z = 1).
The atomic positions are M(a) 0,0,0; 6B(f) $\frac{1}{2}$,
$\frac{1}{2}$, x (x = 0.19538 for our self-consistent
calculation). There are two different B-B distances for intra- and
inter-octahedral B-B bonds.\\ The crystal structure of layered
AlB$_2$-like Mg and Y diborides (space group is
D$^1$$_{6h}$-P6/mmm) is of an entirely different. It contains
graphite-type boron sheets separated by hexagonal close-packed
metal layers. Metal atoms are located at the center of hexagons
formed by boron atoms. The unit cell contains 3 atoms (Z = 1). The
atomic positions are M (a): 0,0,0; 2B (d): $\frac{1}{3}$,
$\frac{2}{3}$, $\frac{1}{2}$ and $ \frac{2}{3}$, $\frac{1}{3}$, $
\frac{1}{2}$. AlB$_2$- type diborides exhibit a strong anisotropy
of B-B bond lengths: the inter-plane distances are considerably
higher than in-plane B-B distances. \\Table 1 lists the lattice
parameters of the borides under study both taken from experiments
\cite{Kuzma} and obtained in our structural relaxation
calculations.\\ The electronic band structures of the
above-mentioned yttrium and magnesium borides were calculated
using codes \cite{Savrasov}. This program employs a scalar
relativistic self-consistent full-potential linear muffin-tin
method (FLMTO) within the local density approximation (LDA)
 \cite{Methfessel} with allowance for correlation and exchange
effects \cite{Perdew1} by using the generalized gradient
approximation (GGA) \cite{Perdew2}. The tetrahedron method was
used to calculate the density of states.

\section{Results and discussion}

\subsection{YB$_{12}$}
Fig. 1 displays the energy bands of YB$_{12}$ as compared with the
hypothetical "dodecaboride" EB$_{12}$ with a vacant Y sublattice.
For EB$_{12}$, the total width of the valence band (VB) is 10.32
eV (without quasi-core B2s-type flat bands located at $\sim$ 14 eV
below the Fermi level). It includes two groups of occupied hybrid
B2s,2p-like bands (A and B) within the intervals -10.99 $\div$
-8.83 and -8.44 $\div$ 0 eV, separated by a gap ($\sim$ 0.4 eV).
The lower bands contain predominately B2s-, and the upper bands -
B2p-type contributions. The latter can be separated into three
groups depending on the intra-atomic bonds in the crystal. There
are two types of bonding B2s,2p states in the spectrum of
EB$_{12}$. The states of the first type (covalent
intra-icosahedral B-B interactions, namely three-center bonds on
the triangular faces of icosahedra) are responsible for the
stabilization of individual B$_{12}$ polyhedra and depend little
on the their packing (B$_{12}$ sublattice structures) and the
interaction of B$_{12}$ between each other (as well as with second
sublattice atoms). They correspond to high k-dispersion bonding
bands located below E$_F$ and are similar  for BB$_{12}$ and
YB$_{12}$, see Fig. 1,2.\\The states of the second group involve
inter-icosahedral interactions. The nonbonding  2s,2p states of
boron atoms surrounding the vacant sites in   B$_{12}$  belong to
third group. The narrow B' and B'' peaks in DOS correspond to it.
.These states form, in particular, a set of very flat partially
occupied bands near the Fermi level (peak B'') with large
effective masses. Consequently, EB$_{12}$ is a metal similar to
$\alpha$-B$_{50}$ \cite{Li}. As distinct from the stable
insulating state of $\alpha$-B$_{12}$ [5-8], where all mentioned
above bands are fully occupied, the "deficiency" of electrons
determining the metal-like properties of the hypothetical
EB$_{12}$ is brought about by structural peculiarities of the
modeling B$_{12}$ crystal. There are "empty spheres" in its bulk
on the place of removed Y atoms. This intrinsic hole region
accumulates some electrons (0.95 e according to our estimates)
leading to partial devastation of the upper boron bands. DOS at
the Fermi level (N(E$_F$) = 6.177 eV/cell) consist of mainly B2p
component ( $\sim$ 96 percentage).\\ The spectrum of EB$_{12}$
contains a wide forbidden band of 1.36 eV (FG, direct transition
in X point) comparable with the FG in $\alpha$ -B$_{12}$ (indirect
$\Gamma$ $\rightarrow$ Z gap of $\sim$ 1.43 - 1.70 eV [5-8]).\\
The main differences in band structures of YB$_{12}$ and EB$_{12}$
are determined by valence yttrium s,p,d-states hybridized with
above mentioned inter-icosahedra and nonbonding B2p states, peak C
Fig. 2. For YB$_{12}$, the total width of the valence band is
12.98 eV, including two groups of fully occupied hybrid B2s,2p
bands of 2.82 and 8.89 eV widths separated by a pseudogap. These
bands shows the significant dispersion near the Fermi level. We
also calculated a hypothetical dodecaboride BB$_{12}$
(isoelectronic with YB$_{12}$), where an yttrium atom is replaced
by an "additional" boron atom, see Fig. 1. It was found that (i)
some s,p- states of the "additional" boron atom are located in the
vicinity of E$_F$; (ii) the system is of a metal-like character
with a rather high density of states at the Fermi level (N(E$_F$)
= 3.034 eV/cell), the main contribution being from the B2p states
( $\sim$ 72 percentage).\\ As noted above, the experiments were
reported recently, in which boron transforms from the usual
non-metal to the SC state at high pressures (above 160 GPa), the
crystal structure of the SC boron remains unknown \cite{Maihion}.
This finding was explained \cite{Maihion} on the basis of the
electron-phonon mechanism in the assumption that under high
pressure  -boron undergoes structural transformations to a simple
fcc B having a metal-like energy spectrum, with N(E$_F$) = 0.154
states/eV cell (lattice constant 2.44 $\AA$) and the dominant
contributions from the B2p states ($\sim$ 60percentage). According
to Table 2, the near-the Fermi level  regions of our hypothetical
systems EB$_{12}$ and BB$_{12}$ have  a similar band structure.
The former system can be interpreted as a model of elemental boron
with a disordered lattice. The latter imitates the role of
inter-icosahedral boron atoms in crystal. This allows us to
surmise that the observed \cite{Eremets} transition of the
rhombohedral $\beta$ -boron to the superconducting state may be
due both to lattice disordering and partial frustration of the
initial icosahedral units, when some boron atoms take
inter-icosahedral positions as a result of high pressure. Such
processes can be more probable than the phase transformation of
rhombohedral  -boron to the fcc structure proposed in
\cite{Papaconstantopoulos} if we take into consideration the high
cohesion properties of elemental boron [5-7].

\subsection{YB$_6$}

Fig. 1 presents the electronic band structure of Y hexaboride. The
10 occupied energy bands is made up by hybrid B2s,p-states
represented the inter- and intra-octahedral B-B bonds. The VB
width in YB6 (without quasi-core B2s band) is about 11.80 eV. The
highest fully occupied bands is due to Bp$_{x,y}$-states  formed
by inter-octahedral interactions. They have strong dispersion
along $\Gamma$ -X direction due to the formation of  hybrid Y-B
bonds. The DOS spectrum has two maxima (A and B, Fig. 2),
corresponding to hybrid states,  formed by  covalent  B B bonds
inside and between B6 clusters. This feature is typical for all
CaB$_6$ like hexaborides [15-17]\\. The partially occupied band
contains a considerable contribution from cation states and has
also a large wave-vector dependence, which reflects the
delocalized character of Yd states forming the bottom of the
conductivity band.\\ It has been indicated that YB$_{12}$ and
YB$_6$ are the conventional low-temperature phonon-mediated BCS
superconductors, review \cite{Buzea}. Thus, the important
parameter is the orbital composition of N(E$_F$). According to our
data, the lowering of the transition temperature (7.1 K (YB$_6$)
$\rightarrow$  4.7 K (YB$_{12}$)) can be explained by a
considerable decrease of contributions of Yd-states to N(E$_F$)
from 0.798 (YB$_6$ $\sim$ 71 percentage) to 0.538 ( YB$_{12}$,
$\sim$ 35 percentage per atom in unit cell, Table 2).

\subsection{YB$_2$ as compared with MgB$_2$}

The electronic structures of layered AlB$_2$-like Mg and Y
diborides are of a completely different, Fig. 3, 4. The
peculiarities of the band structure of the SC MgB$_2$ are due to
the B2p states which form four $\sigma$(2p$_{x,y}$) and two
$\pi$(p$_z$) bands, Fig. 3. The E(k) dependence for B2p$_{x,y}$
and 2p$_z$ bands differs considerably. For B2p$_{x,y}$) like bands
the most pronounced dispersion of E(k) is observed along the
direction k$_{x,y}$ ($\Gamma$ - K of the Brillouin zone (BZ)).
These bands are of the quasi two dimensional (2D) type. They form
a flat zone along k$_z$ ($\Gamma$-A) and reflect the distribution
of $\sigma$(2p$_{x,y}$) states in the boron layers. These states
make a considerable contribution to the N(E$_F$) forming metallic
properties of the diboride. E$_F$ is located in the region of
bonding states and, the conductivity of MgB$_2$ is due to hole
carriers. Mg is ionized, the charge transfer takes place in the
direction Mg $\rightarrow$ B. B2p$_z$-like bands are responsible
for weaker $\pi$(p$_z$) interactions. These 3D-type bands have the
maximum dispersion in the direction k$_z$ ($\Gamma$-A). Mgs,p and
Bs states are admixed to B2p-like bands near the bottom of the VB
and in the conduction band. Therefore the peculiarities of the
electronic properties of MgB$_2$ are associated with the
metal-like 2p states of boron atoms located in plane nets.  These
states determine the DOS in the vicinity of the Fermi level and is
an important condition for superconductivity in MgB$_2$ and
related phases [2,3, 19-21].\\Thus the crucial features of the
band spectrum of MgB$_2$ for its superconducting properties (see
also [2,3,19-21]) are: (i) the position of $\sigma$(p$_{x,y}$)
bands relative to E$_F$ (the presence of hole p$_{x,y}$- states);
(ii) their dispersion in the direction $\Gamma$ - A
($\Delta$E$^{\sigma}$($\Gamma$-A) is determined by the interaction
between metal-boron layers); (iii) the value and orbital
composition of N(E$_F$) (the main contributions from boron
$\sigma$ -states).\\ Let us compare the band structures of MgB$_2$
and YB$_2$. The most obvious consequence of the metal variation
(MgB$_2 \rightarrow$ YB$_2$) is band filling change caused by the
increased number of valence electrons. For YB$_2$, the Fermi level
is shifted towards a pseudogap between bonding and antibonding
states. As a result, the near-E$_F$ spectra of YB$_2$ and MgB$_2$
differ radically. For YB$_2$, (i) $\sigma$(p$_{x,y}$) boron bands
are almost filled and the hole concentration is very small (near
point A of the BZ, Fig. 3); (ii) the covalent d-p metal-boron
bonding increases considerably and the B2p-like bands are shifted
downwards to point K (these interactions are also responsible for
the appearance of pronounced dispersion of bands in the direction
$\Gamma$ $\rightarrow$ - A and for 2D $\rightarrow$ 3D
transformation of the near-E$_F$ states); (iii) the Y4d band along
the $\Gamma$ -M is below E$_F$ and these states
give a large ($\sim$ 59 percentage) contribution to N(E$_F$).\\
The 2D $\rightarrow$ 3D transformation of the near- E$_F$ states
can be also traced by comparing the the Fermi surfaces (FS) of
MgB$_2$ and YB$_2$, Fig. 3. For MgB$_2$, B2p$_{x,y}$ bands form
two hole-like cylindrical Fermi surfaces around the $\Gamma$ - A
direction. The other tubular surfaces come from bonding
(hole-like) and antibonding (electron-like) B2p$_z$ bands. By
contrast, the FS of YB$_2$ consists of hole-like ellipsoids around
the $\Gamma$ - A line and 3D figures with electron-type
conductivity. All these FSs are defined by mixed Y4d-B2p states.
Thus, the absence of superconductivity in YB$_2$ can be accounted
for (see also \cite{Medvedeva}) by the high energy shift of E$_F$,
a considerable increase of Y4d-component in N(E$_F$), and the
absence of  $\sigma$(p$_{x,y}$) hole states at $\Gamma$, Fig. 3.

\section{Conclusions}

We presented the results of  full-potential LMTO band structure
calculations for yttrium dodeca- and hexaborides compared with
layered non-superconducting YB$_2$ and the new "medium-T$_c$" SC
MgB$_2$ diborides.\\
The band structures of boron-rich crystals are determined by the
complicated intraatomic bonds including intra and between boron
polyhedra  (B$_{12}$ for YB$_{12}$ and B$_6$ for YB$_6$) and
direct Y-B bonds. The DOS at the Fermi level for the
low-temperature SC YB$_{12}$ and YB$_6$ has a similar composition
and includes the large contribution of Y4d-states. The lowering of
T$_c$ (YB$_6$ $\rightarrow$ YB$_{12}$) can be explained by a
considerable decrease in main contributions of Yd-states to
N(E$_F$). We performed also band structure calculations for two
hypothetical structures: "dodecaboride" with an empty Y-sublattice
(EB$_{12}$) and the icosahedral phase (BB$_{12}$), which is a
result of the Y $\rightarrow$ B substitution. Both crystals are
metals with high DOS at the Fermi level. We speculate that the
observed \cite{Eremets} transition of the rhombohedral $\beta$
-boron to the superconducting state can be due to both lattice
disordering and partial frustration of the initial icosahedral
units as a result of high pressure. Such processes can be more
energetically favorable than the phase transformation of
rhombohedral $\beta$ -boron to the fcc structure proposed in
\cite{Papaconstantopoulos}.\\ Quite different are the band
structures of layered AlB$_2$-like Mg and Y diborides. They are
determined by intra- and interlayer interactions of plane (Mg,Y)
and boron nets. In contrast to MgB$_2$ , for YB$_2$  the increase
of covalent d-p bonding leads to the downward shift of B2p$_z$
bands, larger dispersion of $\sigma$ bands in the direction
$\Gamma$ - A and 2D $\rightarrow$ 3D  band transformation near-
E$_F$. The most crucial changes are connected with the increase of
electron numbers resulting  in the almost filled
$\sigma$(p$_{x,y}$) boron bands. The hole $\sigma$(p$_{x,y}$)
bands are absent at point and  the density of states at the Fermi
level is mainly defined  by  Y4d states, so these band structure
peculiarities may be considered to be responsible for the absence
of medium-T$_c$ superconductivity in YB$_2$.

\onecolumn

\begin{table}
\caption{Transition temperatures (T$_c$, K), structure type and
lattice constants ($\AA$) YB$_{12}$, YB$_6$, YB$_2$ and MgB$_2$. a
and c are lattice parameters from \cite{Kuzma}, a$^*$ and c$^*$
are lattice parameters from our self-consistent data.}

\begin{center}
\begin{tabular}{|c|c|c|c|c|c|c|}
\hline

Boron & T$_c$ & Structure type& a & c & a$^*$ & c$^*$ \\
      & & (space group)& & & & \\

\hline
 YB$_{12}$ & 4.7 & UB$_{12}$(Fm3m) & 7.5000 &  - & 7.5223 & - \\
\hline
 YB$_6$    & 7.1 & CaB$_6$(Pm3m) & 4.1132   &  - & 4.1554 & - \\
\hline
 YB$_2$    & --- & AlB$_2$(P6/mmm)& 3.3036 & 3.8427& 3.2116 & 4.0080\\

\hline
 MgB$_2$   &$\sim$ 40& AlB$_2$(P6/mmm)& 3.083 & 3.521 & 3.0487 & 3.4664\\

\hline
\end{tabular}
\end{center}
\end{table}

\begin{table}
\caption{Total and site-projected l - decomposed DOS at the Fermi
lavel (state/eV/cell) for borides.}
\par
\begin{center}
\begin{tabular}{|c|c|c|c|c|c|c|}
\hline
 Boride & Total & Ms & Mp & Md & Bs & Bp \\
\hline
 YB$_{12}$ & 1.458 & 0.005 & 0.003 & 0.532 & 0.033 & 0.885\\
\hline
 EB$_{12}$ & 6.177 & 0.240 &       &       & 0.033 & 5.904\\
\hline
 BB$_{12}$ & 3.034 & 0.743 & 0.119 &       & 0.109 & 2.063\\
\hline
 YB$_6$    & 1.130 & 0.017 & 0.020 & 0.798 & 0.001 & 0.294\\
\hline
 YB$_2$    & 1.665 & 0.042 & 0.106 & 0.983 & 0.006 & 0.528\\
\hline
 MgB$_2$   & 0.719 & 0.040 & 0.083 & 0.138 & 0.007 & 0.448\\
\hline
\end{tabular}
\end{center}
\par
\end{table}

\begin{center}
Figures.\\
Fig.1. Band structures of YB$_{12}$ (I), EB$_{12}$ (II), BB$_{12}$
(III) and YB$_6$ (IV).\\
Fig.2. Total and partial DOS (1- s, 2 - p, 3 - d) for YB$_{12}$,
EB$_{12}$ and YB$_6$.\\
Fig.3. Band structure and Fermi surface for YB$_2$ (I) and
MgB$_2$ (II).\\
Fig.4. Total and partial DOS (1-s, 2 - p, 3 - d) for YB$_2$ (I)
and MgB$_2$ (II).\\
\end{center}

\begin{figure}
\begin{center}
\includegraphics[scale=1.5]{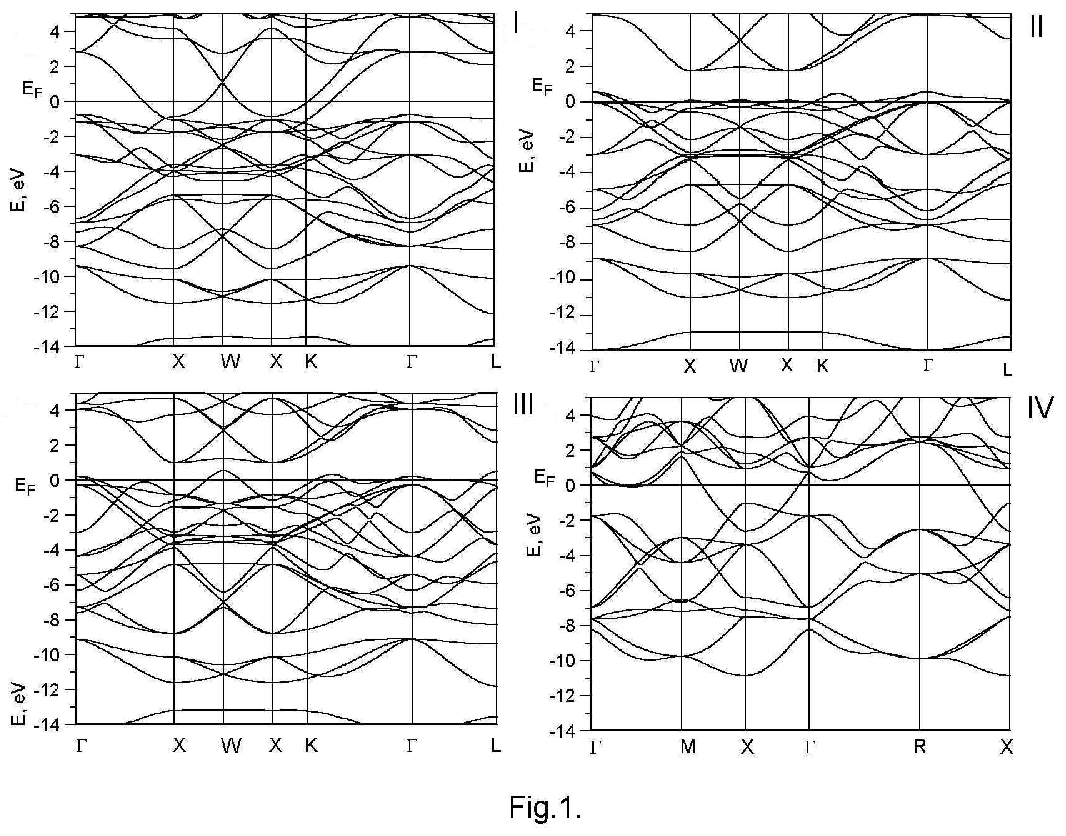}
\end{center}
\end{figure}

\begin{figure}
\begin{center}
\includegraphics[angle = 90, scale=1.5]{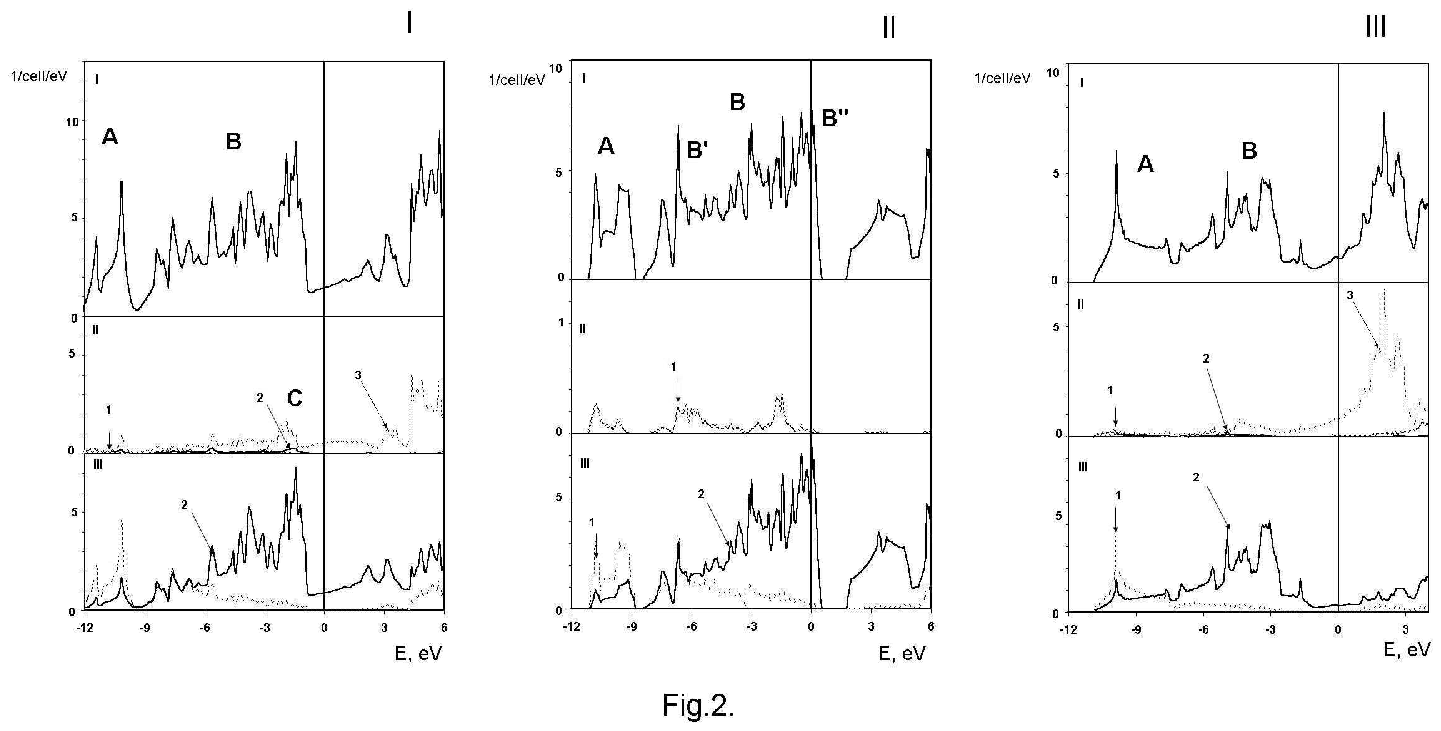}
\end{center}
\end{figure}

\begin{figure}
\begin{center}
\includegraphics[scale=2.5]{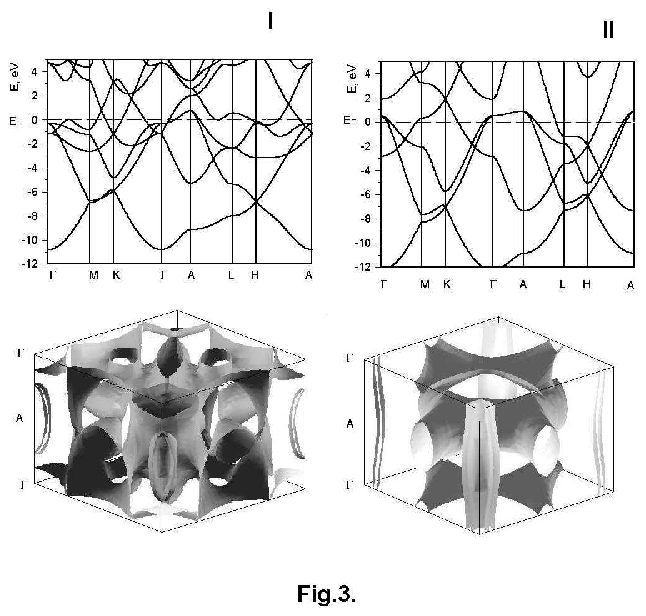}
\end{center}
\end{figure}

\begin{figure}
\begin{center}
\includegraphics[angle = 90, scale=1.75]{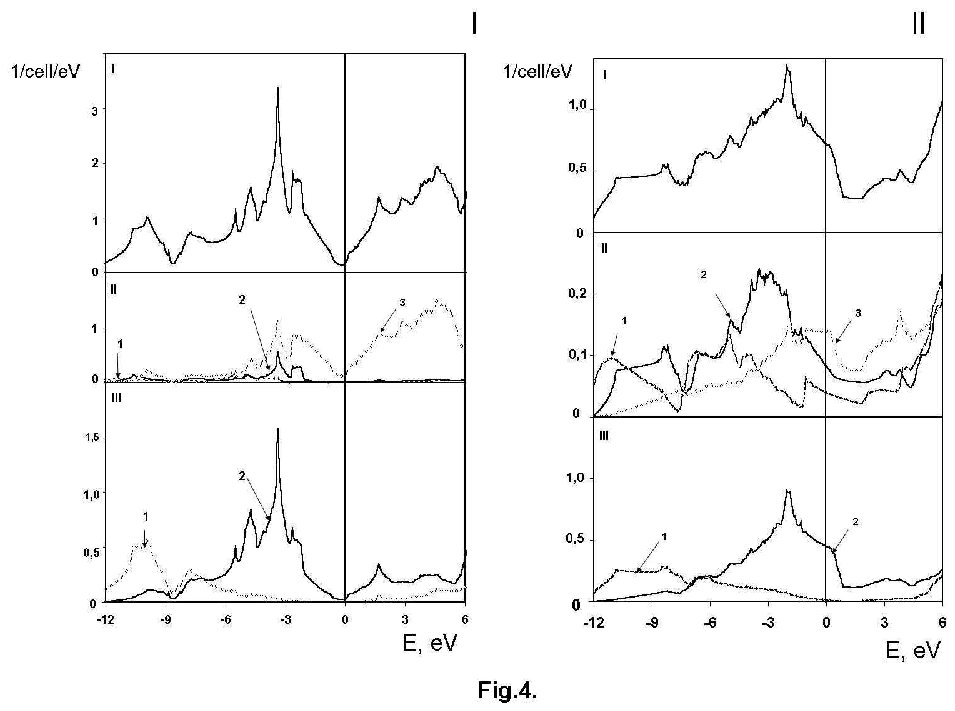}
\end{center}
\end{figure}

\end{document}